\documentstyle[seceq,epsf]{ptptex}
\title{Black Holes and Relativistic Jets}
\author{%       %Use \sc for the family name
R. D. {\sc Blandford}
\footnote{E-mail address: rdb@caltech.edu}
}

\inst{%         %Affiliation, neglected when [addenda] or [errata]
Theoretical Astrophysics, Caltech, Pasadena, CA 91125, USA}

\recdate{%      %Editorial Office will fill in this.
}
\abst{There is strong observational evidence that AGN, Galactic 
X-ray transients and (probably) $\gamma$-ray bursts are associated 
with black holes, and that these sources are able to 
form collimated, ultrarelativistic outflows. There is much  
interest in trying to understand how these prime
movers are able to release energy from accreting gas and their own 
spin energy. 
Electromagnetic field plays a large role in many of the mechanisms 
under active consideration. In this article, several 
of the many possible ``metabolic pathways'' through which 
mass, angular momentum and energy can flow around and away from 
black hole magnetospheres are discussed. Particular importance is attached 
to the interactions between the inflowing disk, the outflowing wind, the
black hole and the jet. Some important unresolved questions are 
identified and it is argued that large scale numerical computation 
will almost certainly be necessary to address them. 
}
\begin{document}
\maketitle
\section{Introduction}
As is well known, one of the first proposals that was made, 
soon after the discovery of quasars in 1963, was that they were powered
by accretion onto massive black holes\cite{rf:zel64}\cite{rf:sal64}. The 
fundamental reason why this proposal was made was that quasars were
known to be prodigiously powerful, with luminosities equivalent 
to hundreds of galaxies, and that up to $\sim0.1c^2\equiv10^{20}$~erg 
g$^{-1}$ of energy 
per unit mass could be released by lowering matter close
to a black hole. This efficiency could be over a hundred times
that traditionally associated with nuclear power. Since this time,
we have also learned about black holes with masses $\sim5-10$~M$_\odot$
in Galactic binary systems, gamma ray bursts and ultra-luminous 
X-ray sources which, with decreased confidence, we also associate 
with black holes, primarily on energetic grounds. 

Simultaneously, remarkable progress was made in understanding the 
theoretical physics of black hole spacetime and in the decade
following the discovery of the Kerr metric in 1963,
the geometrical and kinematic properties of the exterior spacetime 
were comprehensively understood. As a consequence, any physical process that 
could be described in a flat spacetime could also, in principle,
be computed around a spinning black hole. In an almost literal
sense, the stage was set for interpreting the extraordinary observational
discoveries that followed and the fact that we still do not possess 
confident and widely accepted answers to most of the most 
basic questions concerning AGN, X-ray transients and GRBs is that 
we are still wrestling with trying to understand non-relativistic 
processes, most notably those associated with electromagnetic field
and gas dynamics.

Astrophysical black holes form a two parameter family. They can be 
completely characterized by their mass $m$ and specific angular
momentum in geometrical units $a$\cite{rf:mis73}. 
The mass simply provides a scale
of length, time and energy and provides the basis for our discussing stellar 
and massive holes simultaneously. By contrast black holes are almost 
surely spinning quite rapidly as their surroundings are always endowed with
far more specific angular momentum than the maximum that holes can possess
and this second parameter introduces some qualitatively new 
features. (Charged black holes
can also be described and are proving to be extremely valuable to the 
theoretical physicist; however they are thought to discharge almost
immediately under astrophysical conditions. In addition, 
there remains the slim possibility that general relativity 
needs to be modified in these environments and astronomers 
are obligated to test the theory whenever they can. Here, thouhg, I shall
assume that the Kerr metric is all that is required.)

In this article, I would like to present a partial summary 
of some general ways by which gravity, spin and electromagnetic
field can combine to power some of the most  extensively observed high energy
phenomena in astronomy. I shall first give a quick list of some of the 
more pertinent pieces of observational evidence concerning 
black holes. Then I shall discuss the four elements, the disk, the outflow,
the hole and the jet where electromagnetic fields are believed to be
relevant and I shall go on to describe the crucial electromagnetic
interactions between these four elements. 

I should make two apologies in advance.
Firstly, if this article is read by anyone 
who actually attended the meeting, they may notice that
it contains material that I did not discuss. This is because I have lost 
my notes and viewgraphs! Secondly, the literature on this subject has 
become quite large and I apologize for only refering to a few representative
articles which, I hope will provide ongoing access to further research. 
\section{Observed Black Holes}
\subsection{AGN}
Although the first direct evidence for the existence of black holes
was garnered in 1972, the observational database has grown impressively
over the past decade. It now appears that most normal galaxies
with the possible exception of types later than Sbc, harbor massive 
black holes in their nuclei with masses from a million to several billion 
solar masses measured dynamically through the motions of surrounding
stars and gas. It appears that the mass of the hole is correlated
with the properties of the surrounding galaxy, most recently with its 
central velocity dispersion $\sigma$\cite{rf:kor01}
\begin{equation}
M_H\sim1.3\times10^8\left({\sigma\over200{\rm km s}^{-1}}\right)^{3.65}.
\end{equation}
Likewise, it has been possible to measure nearly twenty black hole 
masses in accreting (often transiently) binary systems. We really have no 
idea as to the number of single black holes orbiting our galaxy 
although candidate objects have been reported in microlensing
surveys. More recently, a several compact X-ray sources
have been reported in nearby galaxies radiating well above the Eddington 
limit for stellar mass holes. These are conjectured to be intermediate
holes with masses $>100$~M$_\odot$, possibly formed at very early
cosmological times.  Finally, although the evidence for this is purely
theoretical, most contemporary models for gamma ray bursts
involve either the formation or the augmentation of a stellar mass.

There have been advances towards the important goal of measuring the
second parameter, the spin. Firstly, observations
of some nearby, lower power AGN at X-ray wavelengths have revealed
very broad iron lines\cite{rf:lee99}. 
These are believed to originate in fluorescing
gas on the surface of an accretion disk and the strong 
gravitational plus Doppler redshifts that are observed
indicate that the disk is located very close to the event horizon.
Now it is known that Keplerian orbits are only stable beyond
a radius $r=6m$ for a non-spinning, Schwarzschild hole and disks exterior
to this radius would only produce narrow lines. Therefore, it is argued,
the hole must be rapidly spinning as this allows the disk to exist 
right up to the horizon. Although there are concerns that gas on 
unstable orbits plunging into the black hole might also fluoresce,
the conclusion seems reasonable.  
(Note that this method furnishes the ratio $a/m$
and does not give the mass separately.) 
\subsection{X-ray Binaries}
Another approach, that has been developed in recent years, involves the 
measurement of quasi-periodic oscillations from black hole X-ray 
binaries\cite{rf:str01}.
Here the $\sim1-10$~percent, $Q\sim10-500$ X-ray variations are thought
to come from modes associated with the accretion disk. (To be more precise,
the disk is thought to be the timekeeper; the very hard spectrum
of the emission, strongly suggests that it originates in a coronal
region that is excited by the pulsating disk.) The exact nature 
of these modes remains controversial however. Local waves 
involving horizontal and vertical epicyclic motion, trapped normal modes
that may require reflection from an inner edge and magnetic modes
have all been invoked. What is not at issue, though, is that all
of these mode frequencies depend significantly upon the spin of the 
hole in addition to the mass, and as soon as we can agree upon the
type of wave that we are seeing, it should also be possible to measure
the spin frequency of black holes as well as their mass. In addition,
there is the challenge of using ``diskoseismology'' to test general
relativity, just like helioseismology has been used to test
stellar physics.  
\subsection{Jets}
Undoubtedly the largest body of observational data, that pertains to the 
effects I have been asked to discuss, is associated with relativistic 
jets. These are commonly found to accompany a minority of active galactic
nuclei, especially those associated with elliptical galaxies. The 
ultrarelativistic outflow speeds, have Lorentz factors that are typically 
$\gamma\sim10$ (and which could be much larger), strongly suggested 
that the ouflows are formed quite deep in the potential well. This 
inference has received support from VLBI observations
of M87 which show that collimation takes place within 
$\sim100m$\cite{rf:jun99}.
Observations of this source also demonstrate that the jet power
can apparently exceed the bolometric power of the accreting 
gas and, in general the jet phenomenon has to be seen, on energetic
grounds, as an intrinsic part of the accretion process. In the case
of Cygnus A, for example, the jet power which inflates the two
radio lobes, located well outside the optical image of the galaxy,
must exceed $\sim10^{45}$~erg s$^{-1}$, comparable with the 
bolometric luminosity of a typical quasar\cite{rf:wil00}. 
Jets are increasingly
being found connected to accreting Galactic black holes, the 
so-called Galactic superluminal sources\cite{rf:fen01}. 
As these probably involve
similar processes to those occuring in extragalactic sources, they 
are well worth studying because the cycle times are very short and
and give a far better sampled dataset for studying accretion 
processes.

Our most detailed measurements of jets come from observations
of young stellar objects\cite{rf:bal00}. 
Here, we are able to infer speed and density
as well as observe the strong spatial inhomogeneity. Of course, there is
no black hole, but the presence of a disk, with a measured magnetic
field, as well as a source of free energy associated with the 
relative angular velocity between the inner disk and the more slowly
spinning hole, may well allow it to function in an essentially similar manner 
to the AGNs.
\subsection{Solar Corona}
Another source of indirect, though no less promising, intelligence
is the solar corona\cite{rf:asc00}. 
There has been a revolution in solar physics, 
due largely to observations by YOHKOH, SOHO and TRACE. Observations
of the behavior of the active corona should allow us to divine
the true behavior of cosmic magnetic fields in the bulk. Fundamental
processes like topological rearrangement, reconnection and the formation
of shock waves, can be monitored on a daily basis and ought to be milder 
versions of similar processes taking place above the surface of a more 
rapidly, and differentially, rotating accretion disk. 
\section{The Four Elements: the Hole, the Disk, the Outflow and the Jet}
We have emphasized that there is still no agreement about the fundamentals
of the flow of mass, angular momentum and energy around black holes.
Part of the problem is that these mechanisms may vary considerably from one 
class of objects to the next. After all, we need to explain why 
broad absorption line quasars differ from powerful
radio galaxies and so on. It 
may be helpful to take a ``systems engineering'' approach and suppose
that the four essential elements are the hole, 
(surely spinning), the inflow (almost certainly 
in the form of a disk or a torus), the outflow (a wind derived from the 
disk) and the relativistic jet (that
derives from the region around the black hole). The difficulty
of the problem is that there are interactions between all pairs
of elements and most of the controversy 
comes about in assessing the character and strength of these interactions. 
However let us first consider the individual elements (considered 
whimsically, using an appeal to Greek philosophy) in turn.
\subsection{``Earth''- the Hole}
As mentioned above black hole
astrophysics is almost exclusively studied in the context of the 
Kerr metric. This is pragmatic because it is much harder to 
amass evidence from observing black holes that general relativity 
failed in this regime than it is to interpret what we see 
assuming that it is correct.
\subsubsection{Reducible Mass}
As is well known, a non-rotating hole introduces an event horizon 
with radius $r_+=2m$ ($G=c=1$) into 
the spacetime across which radiation and plasma can only cross inwards. 
In order to describe the spacetime surrounding
the hole, we must introduce a coordinate system. The easiest one to use,
Schwarzschild coordinates, has a pathology at the horizon and kinematic
thought experiments have to be interpreted carefully, often with the 
aid of alternative, non-singular coordinate systems\cite{rf:mis73}.
For example, particles remain outside the horizon 
after the Schwarzschild time coordinate -- essentially time measured
at infinity --  has advanced by an infinite amount. For this reason,
black holes were once called ``frozen stars''. However, this 
is not a useful way to view the physics. According to a clock
attached to an infalling particle, it only takes a finite proper time
to cross the horizon.

Rotating black holes exhibit the effects of spin. The radius of 
the event 
horizon in Boyer-Lindquist coordinates - the 
generalization of Schwarzschild coordinates - changes to 
\begin{equation}
r_+=m+(m^2-a^2)^{1/2}
\end{equation}
where $a$ measures the angular momentum per unit mass.
In addition, the spacetime is dragged
in the direction of rotation. In practice, what this means
is that a material particle, that orbits the hole at fixed latitude
and radius, must rotate with respect to infinity when it is within 
a region called the ergosphere which is
defined by $r<r_{{\rm ergo}}=m+(m^2-a^2\cos^2\theta)^{1/2}$.
To be more precise, the angular velocity of these 
particles (which require non-gravitational forces to keep them
on their trajectories) must lie between two limits      
\begin{equation}
\Omega_{{\rm min}}<\Omega<\Omega_{{\rm max}}
\end{equation}
and $\Omega_{{\rm min}}>0$ within the ergosphere.
In the limit as $r\rightarrow r_+$, $\Omega_{{\rm min}}
\rightarrow\Omega_{{\rm max}}
\rightarrow\Omega_H=a/(r_++a^2)$ which is defined to be the angular 
frequency of the hole. (Of course this angular frequency cannot be
measured directly as the spacetime is axisymmetric. However,
it can be measured indirectly using gyroscope precession etc.)

One of the most perceptive theoretical discoveries that was made 
about black holes was that, when they spin, a fraction of their
mass could be ascribed to rotational energy and was, in principle,
extractable\cite{rf:pen65}. This is most convincingly demonstrated
by observing that there exist
orbits of test particles with negative total energy, (including their
rest mass), within the ergosphere. This implies that if, for example,
a cloud of plasma is attached to magnetic field lines, anchored
at large distance and this magnetic field drags the 
cloud backwards relative to the rotation of the hole so that it is placed
onto a negative energy orbit, then the work that was performed in placing 
the cloud onto this orbit can be thought of as energy that has been 
extracted from the spin of the black hole\cite{rf:ruf75}. 
After the cloud has crossed
the event horizon, the gravitational mass of the hole will decrease
by the mass equivalent of the work that was performed. (Note that this is 
a purely classical process, in contrast to Hawking radiation.)
Obviously, this requires that
we define energy (and angular momentum) in a consistent fashion,
and this happens naturally when one uses the full machinery of general 
relativity. 

When we generalize this idea, we find that associated with  
a black hole of mass $m$ is an irreducible mass $m_0$ related to $m$
through 
\begin{equation}
m=m_0(1-\beta^2)^{-1/2}
\end{equation}
where $\beta=a/2m_0=(a\Omega_H)^{1/2}<2^{-1/2}$. The reducible 
mass, $m-m_0$ is available for powering high energy emission.
\subsection{``Water'' - the Disk}
The accreting gas almost certainly has sufficient angular momentum
to form a disk or a torus, if the pressure support is substantial.
Several different modes of accretion have been proposed,
dependent upon the rate of gas supply
and possible other factors, like the spin of the hole or the stellar 
environment (binary companion, massive star envelope, dense star cluster 
and so on). 
\subsubsection{Thin Disks}
In the simplest and most standard type of accretion disk,
the flow is supposed to be stationary and efficiently radiative.
One peculiarity of thin, radiative disks, that has consequences for what 
follows, is that when the torque, $G(r)$, that drives the inflow is local, 
then three times as much energy is radiated from 
an annular ring (well removed from the disk boundaries and assuming
a steady state) than is released 
by the infalling gas. The reason for this is that the torque 
that the inner disk exerts on the outer disk and which balances
the inflow of angular momentum $\dot M(mr)^{1/2}$ (where $\dot M$
is the accretion rate) does work at a rate
$G\Omega$ which overcompensates the inflow of binding energy. The 
divergence of the energy flux is $-Gd\Omega/dr$ which equals
three times the rate of release of binding energy. Energy 
is conserved overall as the boundary conditions ensure that there is 
a shortfall in the energy radiated close to the inner boundary of the disk.
\subsubsection{Adiabatic Accretion}
At sufficiently low accretion
rates the energy that is created by viscous dissipation 
may be almost completely taken up by the ions. (There are some 
plasma physical arguments that this is indeed what happens.) Under these 
circumstances, the gas cannot cool and maintains a pressure
scale height comparable with its radius. However, this cannot happen 
too close to the symmetry axis where horizontal, centrifugal force 
cannot oppose the vertical pull of gravity and a funnel is expected to
form.  There should still be a
viscous torque carrying angular momentum outward. It turns 
out that the energy dissipated by this torque is sufficient to unbind
the gas. It has been proposed\cite{rf:bla99}
that, under these conditions, inflow into the hole
should be accompanied by prodigious mass loss from larger radii 
driven by the binding energy released by gas as it 
crosses the horizon. (Under these
circumstances, the gas is likely to be convective.) The outflow is most 
commonly argued to be magnetically driven\cite{rf:kra99}. 

There are, at least, two 
alternative views. Under the ``ADAF'' prescription\cite{rf:nar94}, the flow is 
conservative and the released energy 
is advected inward across the horizon in a quasi-spherical
inflow. Under the accretion torus model\cite{rf:pac98}, 
the inner surface of the disk has a 
very small binding energy from which gas falls ballistically onto the hole
and the overall radiative efficiency is very small.

These matters have become important in interpreting recent Chandra
observations of massive black holes in dormant galactic nuclei. These 
show that the holes are spectacularly underluminous, especially in relation 
to the estimated rate of mass supply. The best studied case is surely 
our Galactic center, where we know the hole mass to be $2.6\times
10^6$~M$_\odot$ and the rate of mass supply is roughly
$\sim10^{22}$~g s$^{-1}$ while the bolometric luminosity is 
$\sim10^{36}$~erg s$^{-1}$\cite{rf:bag01}. 
The detailed X-ray observations suggest 
that the emission is due to nonthermal emission 
from close to the hole.

There is another limit where these considerations may be very important.
This is when the rate of mass supply is much greater than the Eddington
rate and there is no difficulty in  emitting radiation. The problem is in 
allowing these photons to escape from the Thomson thick inflowing gas.
Again, it is likely that the inflow is effectively 
adiabatic, (though recent intriguing suggestions as to how radiation
may escape through magnetized channels challenge this view\cite{rf:beg01}). 
However, if we suppose that super-Eddington accretion is adiabatic, then
it too may be accompanied by prodigious mass loss. Perhaps this is what
is happening in broad absorption line quasars (BALQ).
\subsection{``Air'' - the Outflow}
Having made the case that accretion at both high and low rates
is accompanied by powerful winds, it is worth thinking about the 
implications. We already know that these winds are a prominent feature 
of accretion onto young stellar objects and they may well carry off
much of the angular momentum and even much of the liberated energy.

As we discuss further below, these outflows may or may not collimate.
If they do, then they become jets and are likely to be able to propagate 
to large distances from their sources as is observed in radio galaxies,
quasars and, especially young stellar objects. If they do not, then they 
are likely to become supersonic and pass through a strong shock when
their momentum flux falls to match the ambient pressure. These shocks
could be sources of nonthermal emission as they are likely to accelerate
relativistic electrons. In addition they may have an important role
in shaping the accretion flow.

The BALQ outflows are particularly interesting 
as it is highly likely that they are subject to additional radiative 
acceleration due to the momentum flux carried by the ``absorbed''
(actually scattered) resonance 
line photons\cite{rf:wey01}. Naive arguments hold that 
the flow is highly inhomogeneous and composed of clouds with dimension
given by the local pressure scale height $\sim r/M^2\sim10^{10}$~cm, where
$M\sim30000{\rm km s}^{-1}/10{\rm km s}^{-1}$ is the Mach number. It is 
possible that these clouds might be the result of a radiation-driven 
instability. (These remarks might also be relevant to SS433.)
\subsection{``Fire'' - the Jet}
Finally, there is the region extending from around the 
black hole and the inner edge of 
the accretion disk and extending out along
the symmetry axis to form a jet. We presume
that the plasma density will be very low and particle acceleration 
is very efficient. 
This is also a region where the optical depth to Thomson scattering is likely
to be quite low and bright ionizing radiation may be able to escape to form
an ionizing cone. 

Following successful detections by the EGRET telescope on 
CGRO, it has been possible to trace relativistic jets down 
to what are inferred to be even smaller radii than the ``smoke''
observed using VLBI. There is almost universal agreement that the 
gamma rays that are observed are produced by inverse Compton scattering
by GeV-TeV electrons in a collimated relativistic outflow. However
there has been controversy concerning the next level of description.
In particular, it is not known whether or not the plasma comprises 
electrons and positrons or if it is a normal electron-ion plasma. 
The former option seems to be the more likely on the same
general grounds that lead one to this conclusion in the case
of pulsars. However it is quite hard to give a compelling 
observational demonstration. One approach is to try 
to estimate the power carried by the jets by observing them 
on large scales and then comparing this with the power carried 
by relativistic electrons derived on the basis of their radio emission.

This can be made a bit more precise in a 
self-absorbed synchrotron source, close to the inverse Compton limit.
The energies of the emitting electrons at the radio photosphere must be 
$\sim300$~MeV while lower energy electrons are essentially 
invisible. However, if they have proton partners, 
the lower cut off in the electron distribution
function must exceed $\sim30$~MeV, otherwise there would 
typically be too much jet kinetic energy,
carried by the protons. Alternatively, if the partners are positrons, 
the pair distribution function
can extend down to mildly relativistic energy without producing large 
jet powers.  
Similar conclusions are drawn from 
observations of strong linear polarization 
which limits the internal Faraday rotation 
within the source. Finally the detection 
of persistent circular polarisation 
has been used to rule in favor of a pair plasma,
on the grounds that the frequency dependence is better fit 
by Faraday conversion
in a pair plasma than intrinsic circular polarisation in a proton 
plasma\cite{rf:war98}.
Note that, if the radiation mechanism is coherent cyclotron 
radiation, then the fact that 
the polarisation is not even stronger than observed may be an indication 
that there is a pair plasma present.

One thing that is very clear is that jets cannot only comprise    
pairs close to the black hole.
This is because pairs are subject to 
strong radiative drag and their outflow speeds would be limited 
to no more than mildly relativistic values. If protons are 
precluded then this suggests that the jet momentum is carried
by electromagnetic field, at least initially. This is really not a 
very unusual conclusion. It is just what is thought to happen in 
the case of radio pulsars, like the one in the Crab Nebula.

Another controversy concerns the origin of the soft photons that are
Compton-scattered by relativistic jets as $\gamma$-rays. 
One idea is that the soft photons originate
within the jet as synchrotron emission, the so-called synchrotron
self-Compton process. The other is that they originate outside the jet and 
are scattered into it. The resolution of this controversy appears to be that
external scattering dominates in the most powerful sources and the 
self-Compton process is most important in weaker sources.

Many radio jets are also observed using HST at optical wavelengths
and, quite recently at X-ray energies using Chandra. The emission 
mechanisms seem to be quite varied. In some cases, like the X-ray knots
of M87, it appears that synchrotron emission by $>3$~TeV electrons are 
responsible for the emission. This places impressive demands on the 
acceleration mechanism within the jet. In other cases the emission is 
attributable to inverse Compton scattering of internal, external or microwave
background photons\cite{rf:pad01}.
\begin{figure}
\epsfxsize=8cm
\centerline{\epsfbox{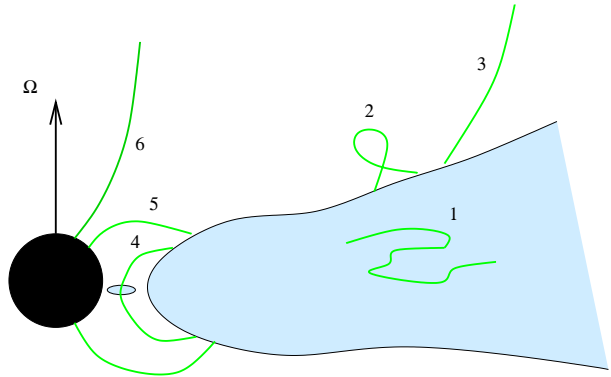}}
\caption{Different types of magnetic field line discussed in the text. 
1. The interior torque is contributed by magnetic field amplified through 
the magnetorotational instability. 2. Short loops of toroidal field
will energise the disk corona, through having their footpoints being twisted
in opposite senses and creating small scale flares. 3. Open field lines
that connect the disk to the outflow may drive a hydromagnetic wind. 
Loops of field from the inner disk that connect to plunging gas (4.)  or the 
event horizon (5.) of the hole can
effectively remove energy from the hole. 
6. Open field lines that cross the event horizon can power a relativistic 
jet, which may be collimated by and possibly decelerated by the 
outflow.}
\label{fig:1}
\end{figure}
\section{The Fifth Element - Magnetic Field}
I now turn to the {\it interactions} between these four elements (Fig. 1). 
(This is, after all, a meeting to celebrate the science of Professor Yukawa!)
As I have
emphasized, these are likely to be quite varied and complex. However they
have the common feature that magnetic fields seem likely to be 
playing a major role. From a theoretical perspective, most attention
has been paid to understanding the ``magnetorotational instability'' (MRI)
\cite{rf:bal98}.
This is a dynamical instability which, in its simplest form 
has a weak, vertical magnetic field becoming unstable with a growth 
time of order the orbital period. What it
is really saying is that weakly magnetized, conducting disks are not viable.
The only possibility is that disks be imbued with magnetic field of 
nonlinear strength which is responsible for the internal torque.
The saturation field amplitude is determined by a balance between 
nonlinear growth, dissipative processes like reconnection 
and buoyant escape and can really only be estimated through
careful, three dimensional, numerical simulations which are just now
becoming possible. 
\subsection{Disk-Outflow Connection}
If the sun and the simulations, are any guide, then magnetic 
field that is growing in the disk will also escape into the corona.
We expect coronal arches, as well as larger scale magnetic 
structures to be quite common and to be regenerated on an orbital 
timescale. If the footpoints of an arch are at different orbital radii
in the disk then they will separate tangentially in a single period. Field
lines will be stretched across the disk surface and will 
quickly be forced to reconnect. I therefore expect that most 
of the magnetic field that connects the disk to the corona 
will be toroidal and that most of the magnetic
flux is ``open'' and escapes into the outflow. However,  differential 
rotation acts in opposite senses on the footpoints of 
a single flux loop and this will cause the loop to twist 
and probably to undergo
some topological re-arrangement, just like a speeded-up 
solar flare. This provides one mechanism for heating a disk corona 
and perhaps for driving an outflow through thermal heating. Although these
tangential flux loops may be regenerated by buoyancy, they will
also be stretched and pulled back into the disk. 
Alternatively, it has
been proposed that they will form an inverse cascade, creating larger and 
larger loops\cite{rf:tou96}. 

Magnetized disks can also launch supersonic outflows
\cite{rf:bla00}.The disk may be threaded by a vertical 
field that is unidirectional, at least over an octave of radius. Differential
rotation will also make the field approximately axisymmetric. Not only
will this create a very effective coupling between the disk and the outflow
but it can also facilitate the outflow by launching it either  
centrifugally or through the magnetic pressure associated with 
coiled up toroidal field. The outflow may be neither stable,
nor even stationary. Again, if the field 
filaments into individual flux tubes, then twisting of the footpoints
may lead to additional dissipation. One criticism that has been 
leveled against this mechanism is 
that any large scale magnetic flux threading 
the disk will migrate outward faster than it is advected inward if the 
effective magnetic Prandtl number is of order unity.

As mentioned above, a strong vertical field will not only
drive an outflow but may even have implications for the evolution
of the disk. It is possible
that the the dominant torque acting on disk be due to this large scale 
global magnetic field as oppose to the internal torque due to the 
nonlinear MRI. This torque removes energy and angular momentum 
in the ratio $\Omega$ just as required by energy and angular 
momentum conservation in the disk
and so there need be no dissipation  in the disk
associated with an external torque
of this type.

If magnetic winds exist,
then they are likely to be strongly collimating as 
demonstrated by axisymmetric numerical simulations\cite{rf:kra99}. 
The basic collimating
action is due to magnetic field lines that are wrapped around the symmetry 
axis and which exert a tension leading to a radial force density.
\subsection{Disk-Hole Connection}
There is also a possible magnetic connection between the disk and the hole
\cite{rf:hir92}\cite{rf:bla99a}
\cite{rf:gam99}\cite{rf:ago00}\cite{rf:lix00}\cite{rf:van01}.
Again there are several possibilities\cite{rf:cam86}.
If the field is confined to the disk plane and the hole is spinning 
rapidly, then gas that falls in from the inner disk edge,
presumably located close to the marginally stable particle orbit, will slide
along magnetic field lines and allow magnetic energy flux to be transmitted
outward in the form of a torque, thereby increasing the energy per unit mass 
that can be dissipated from the disk from that associated with the 
binding energy of the marginal stable orbit.

However, magnetic field is unlikely to be confined to the equatorial
plane and, besides, this is where reconnection should be very actively
disconnecting the inflow from the disk\cite{rf:arm01}. 
The field lines that leave 
the surface of the inner disk, go up to high latitude and then, connect 
with either the gas plunging into the hole or the event horizon of the 
black hole are likely to be more significant and they, too, can extract 
energy and angular momentum. They have the further advantage that they are 
likely to propagate in a region where the Alfv\'en speed is large 
and there 
is causal contact between the inflow and the disk from closer to the 
hole. 
\subsubsection{Black Hole Magnetosphere}
In fact it is quite likely that, except in the region close to the 
disk or the infalling matter, the Alfv\'en speed will become relativistic
which may allow substantial energy to be extracted from the black hole itself.
\cite{rf:bla77}\cite{rf:tho86}\cite{rf:phi83}\cite{rf:mac84}\cite{rf:sue85}
\cite{rf:zha89}.
Before we discuss this, though, let us make two key points
adopting, for illustration, a typical $10^8$~M$_\odot$ hole 
accreting $\sim10^{-2}$~M$_\odot$~yr$^{-1}$. Firstly, if we assume
a fairly thick disk and a viscosity parameter $\alpha\sim0.01$, 
then the gas density in the disk will be $\sim10^{-11}$~g cm$^{-3}$
and the pressure $\sim10^{8}$~dyne cm$^{-2}$. This can support a field 
of strength $\sim10^4$~G. It is important to realize that it is necessary
for there to be a heavy disk present to contain the field. The 
electrical currents that act as a source of this field are external to the 
hole and the stresses that the field exerts on them are matched 
by pressure and gravity. Secondly, the gas density in the
high latitude region around the hole, 
which we call the black hole magnetosphere, although likely to be quite low 
is not going to vanish altogether. This is important because only a tiny 
density of plasma, in this case $\sim10^{-25}$g cm$^{-3}$ is needed 
to supply enough charge to short out the 
potential differences along the magnetic field lines. It is hard to imagine
that plasma could be excluded from the magnetosphere so efficiently
that $\vec E\cdot\vec B$ is not effectively zero. 
Cross-field diffusion and pair production happen all too readily. Conversely,
provided that
the magnetospheric density is $<<10^{-14}$~g cm$^{-3}$ then the 
formal Alfv\'en speed will be ultrarelativistic so that the conditions are 
effectively electromagnetic and the matter will have almost no dynamical
role. It is not clear if and when this second condition is satisfied, but 
it does not seem unreasonable to hypothesize that black hole magnetospheres
are, in this regard similar to pulsar magnetospheres. The work function of a
classical event horizon is infinite! If so, then 
electromagnetic field in the magnetosphere must be relativistically 
force-free\cite{rf:bla76}.
That is to say, 
\begin{equation}
\rho\vec E+\vec j\times\vec B=0.
\end{equation}
In addition, we also require that the Lorentz invariant $B^2-E^2>0$.
The relativistic force-free condition is mathematically equivalent to 
relativistic MHD, where it is assumed that the
electric field vanishes in the center of momentum frame, of the plasma,
moving with velocity $\vec v$ so that $\vec E+\vec v\times\vec B=0$, 
in the limit that the inertia of 
the plasma can be ignored.
 
These two points cast into
doubt some schemes for extracting energy from strongly charged black holes 
and magnetospheres where the electromagnetic field
is supposed to satisfy the Einstein-Maxwell equations {\it
in vacuo}\cite{rf:van01a}.
\subsubsection{A Digression on Pulsars and Causality}
In order to explain, in more detail, how this energy extraction 
is thought to work under electromagnetic conditions, it may be  
helpful to start with an axisymmetric model of a pulsar\cite{rf:con99}. 
The model is simple
but known to be incomplete -- at the very least, pulsars are not
axisymmetric. Nonetheless, it is helpful for explaining some 
principles. Under this model, we have 
a magnetic field that is frozen into the spinning neutron star and we can 
think of the field lines as moving with the same angular velocity 
$\vec\Omega$ as that
of the star, provided that, as we have asserted above, $\vec E\cdot\vec B=0$.
What we mean by this statement is that the electric field vanishes
in any local Lorentz
frame moving with speed $\vec v=\kappa\vec B+\vec\Omega\times\vec r$,
where $\kappa$ is only limited by the requirement $v<c$. 
In the inertial frame, the electric
field has a divergence which is satisfied by a charge density and currents
flow along the magnetic field, driven by relatively small residual and 
quite possibly transient electric field. The poloidal component of the current
is the source for a toroidal magnetic field whereas the electric field
$\vec E=-(\vec\Omega\times\vec r)\times\vec B$ is poloidal.  
There is a surface,
known as the light cylinder, with cylindrical radius $c/\Omega$, beyond
which plasma, which is tied to the moving magnetic field lines, must move 
outward. The combination
of the poloidal electric field and the toroidal magnetic field leads
to a radial Poynting flux that carries energy away from the star at the 
expense of its rotational kinetic energy; the combination of the poloidal 
magnetic field and the poloidal electric field carry away angular momentum
and the ratio of the two quantities is $dE/dL=\Omega$ which is identical to
the ratio of the energy to the angular momentum lost from the pulsar.
The currents, which carry no net charge away from the star, complete
their circuit at large distance from the pulsar, at least in the model.
In order to do so, there must be either dissipation 
or inertial effects which allows electromagnetic
energy to be converted into kinetic energy or
heat. Most plausibly this would happen at a 
distant shock front, which we call the load.

There is one more point that should be made here, and it is important for 
what follows. This concerns the nature of the small amplitude,
short wavelength wave modes in this system. There are two types of
wave in the rest frame of a plasma 
where $\vec E\cdot\vec B=0$ and particle inertia is 
ignorable. One is a fast mode with $\delta\vec E\propto\vec k\times
\vec B$ which is indistinguishable from a vacuum electromagnetic wave;
the other is an intermediate (or Alfv\'en)
mode, with $\delta\vec B\propto\vec k\times
\vec B$, which travels with phase speed $\hat{\vec k}\cdot\hat{\vec B}$
along $\vec k$ and group velocity $1$ along $\vec B$. 

Let us look at
these modes in the non-rotating, global inertial frame. What we find is that,
if we restrict attention to axisymmetric modes, 
then the fast mode has a toroidal magnetic perturbation and the intermediate
mode has a poloidal magnetic perturbation. This means that 
information about the toroidal magnetic field (and, correspondingly, 
the poloidal current, can propagate inward at 
effectively the speed of light across magnetic field lines even beyond 
the light cylinder. In other words, if we 
change the conditions in the outer dissipation region -- 
for example by changing the resistance in the load in the
electrical circuit -- then this will eventually
react back upon the current flowing through the pulsar 
and change the Poynting flux. However, it will not change the EMF 
appreciably. By contrast, if we consider the intermediate
mode, then the disturbances are constrained to flow along the magnetic 
field direction and can only propagate inward within the light cylinder.
These disturbances essentially carry information about the 
toroidal current and the poloidal 
magnetic field. 

I am belaboring these points in what, I hope, is a non-controversial 
context, because essentially the opposite conclusion has been 
drawn in a recent book by Punsly\cite{rf:pun01} 
(and earlier references cited therein).
There are many points of disagreement but one key difference 
is that Punsly argues that the intermediate mode propagates information about
the global charge and current density 
because they involve non-zero perturbations
to the current density. 
I believe this argument to be incorrect. Just as sound waves do not 
physically transport mass despite having a density fluctuation,
intermediate waves do not physically transport charge. 
Actually, in making the force-free approximation, Eq.~(4.2),
we are implicitly
saying that, although the electromagnetic field evolves according 
to Maxwell's equations, ($\partial\vec B/\partial t=-\nabla\times\vec E$,
$\partial\vec E/\partial t=\nabla\times\vec B-\mu_0\vec j$,
$\partial\rho/\partial t=-\nabla\cdot\vec j$), 
there is no corresponding evolution equation
for the current density which must be determined by regions where inertia
or dissipation are important.
The use of such relations is tantamount to saying that the timescale 
on which charge adjusts locally to the electromagnetic field
is very short compared with the time it takes light to cross 
the circuit. In this way, an 
axisymmetric pulsar can act as a battery with negligible internal resistance
and can drive current around a circuit in a manner that is 
eventually responsive 
to changes in the physical conditions in the load,
well beyond the light cylinder.
\subsubsection{Extraction of Energy from the Hole by the Disk}
Returning to the matter at hand, the field lines that connect the inner
disk to the gas plunging into the hole, may be able to transport 
energy and angular momentum outward in a time-dependent manner. The 
field will quickly be dragged into the hole while its footpoints will 
be (temporarily) anchored in the disk and will rotate with the disk angular
velocity $\Omega$. Now, it is probably the 
case that the angular velocity of the 
black hole exceeds that of all of the disk. (If the disk terminates at the 
marginally stable orbit, then this is true for $\Omega_H>0.093/m$.) 
Under these circumstances, the 
Einstein-Maxwell (plus relativistic force free)
equations can be solved subject to suitable boundary 
conditions at the horizon (essentially that the electromagnetic field
as measured by an infalling observer remain finite). It is found that 
the magnetic field will trail the hole 
and that, if we adopt axisymmetry, a slender equatorial
ring, threaded by flux $\Phi$, will experience
a torque 
\begin{equation}
G={I\Phi\over2\pi}={(\Omega_H-\Omega)\Phi^2\over4\pi^2\Delta R_H}
\end{equation}
where $I$ is the current circulating through the ring and into
and out of the hole. In using this equation we need to 
know that 
the effective resistance of the horizon is effectively
the impedance of free space, multiplied by geometrical
factors, $\Delta R_H\sim60\Delta\theta/
\sin\theta\Omega$. Also, note that current can flow through the hole simply
by having positive charges preferentially cross the horizon at high 
latitude and negative charges cross it at low latitude or {\it vice versa}.

This torque will do mechanical work on the disk at a rate $G\Omega$ and 
the energy transfer will ultimately be dissipated by viscous processes 
in the accretion disk in the form of heat. If we consider the causal structure
of this electromagnetic configuration, then there will be 
an inner light surface, the counterpart of the pulsar light cylinder,
within which intermediate waves cannot propagate outwards and
there will be a fast mode surface, very 
close indeed to the horizon\cite{rf:phi83}\cite{rf:van01}. 
The details of what happens
close to the horizon are, as usual, unimportant as they
are redshifted away. This electrodynamic configuration is, in no essential 
respect, different from what happens if we connect
the flux lines to gas endowed with substantial resisitivity
orbiting in the equatorial plane between
the disk and the horizon\cite{rf:bes01}.

Note, though, that the disk is actually extracting energy from the hole 
rather than the other way around. To see 
why this is possible, we must return to the properties of spinning holes.
When we work in Boyer-Lindquist coordinates, we are exploiting the fact 
that the spacetime is stationary and axisymmetric and the metric is
independent of time and azimuth. 
Associated with these two symmetries, are two conservation laws -  those of 
energy and angular momentum. If we look at the electromagnetic 
Poynting flux in the Boyer-Lindquist frame, then the 
energy and angular 
momentum fluxes are conserved from the horizon outward, despite the fact 
that a physical observer within the ergosphere
must move with respect to this coordinate system. If we transform 
to a frame in which physical observers orbit with angular velocity 
lying in $[\Omega_{{\rm min}},\Omega_{{\rm max}}]$ 
then we find that the energy and angular momentum fluxes
must be transformed and the energy flux must change sign close to the horizon. 
From the point of view of physical observers, the hole is a sink not a source
of energy. The reason for this strange behavior is intrinsicically 
relativistic
and can be traced to the existence of the component $g_{0\phi}$ in the metric
tensor. The hole's spin is communicated to the
world exterior to its event horizon not through outwardly 
propagating electromagnetic disturbances emanating
from close to the event horizon but through the 
metric tensor. It is really the spacetime around the hole, not the event
horizon, from which the energy is being extracted. 
\subsubsection{Quasi Periodic Oscillations} 
How much of the above actually happens and how important it is quantitatively
depend upon the details of the gas flow and the magnetic field configuration
that develops. However, if the hole-disk connection is important, then it does
open up the possibility that quasi-periodic oscillations might
be excited by interaction with the hole\cite{rf:bla99a}.
\subsection{Hole-Jet Connection}
Having explained how to connect a disk to a hole and extract energy,
we can now explain how to connect the 
hole directly to a jet\cite{rf:bla77}\cite{rf:lee00}. 
(We do, however, still need a disk to retain the poloidal magnetic field.)
Let us do this in two stages. Suppose that magnetic field lines 
which cross the horizon are frozen into
a highly conducting ring orbiting the hole at high latitude at a radius 
beyond the ingoing light cylinder. Now make charges available at the 
inner surface of the ring and allow current to flow around the 
circuit and energy to be extracted from the hole,
just as we did with the disk. Next make charges 
available on the other side of the ring and allow it to behave like an
axisymmetric pulsar
and lose energy to a distant load.  
In general, the rate of energy (and angular momentum)
gain from the hole at the interior surface of the ring will not balance the
loss from the exterior surface to the load. However, if 
we adjust the angular frequency
of the ring and the magnetic field lines which are frozen into 
it to a suitable 
compromise, between that of the hole and the load, determined by 
a simple circuit analysis, 
\begin{equation}
\Omega={\Omega_H\Delta R_L+\Omega_L\Delta R_H\over\Delta R_L+\Delta R_H}
\end{equation}
where $\Omega_{H,L},\Delta R_{L,H}$ are the effective 
angular velocity and resistance of the load and the hole respectively, 
then we should be able to achieve balance. Furthermore,
this equilibrium should 
be stable. At this point, we can remove the conductor. Provided that
we are still able to make charges available in the region between the two 
light surfaces, as we have argued will be the case, then currents can flow 
all the way around a circuit spanning the event horizon and the load. Energy
will be lost by the hole and dissipated in the load. In fact there 
is generally some dissipation in the hole and although its gravitational
mass may decrease, its irreducible mass will increase. 

This is the basis of the proposal that spinning black holes power jets
and, possibly, gamma ray bursts. The power is created as a large scale 
Poynting flux or equivalently as a battery-driven current flowing around an 
electrical circuit. (For the example quoted, the power evaluates 
to $\sim10^{43}$~erg s$^{-1}$ and the order of magnitude estimate of the 
EMF and the current are $\sim10^{19}$~V and $\sim10^{17}$~A.) This neatly 
avoids the problem of catastrophic radiative drag close to the jet origin.
and allows the terminal jet Lorentz factors to be very large
as observations indicate is the case.

One objection to this model is that the flux threading the black
hole may be small
\cite{rf:bla77}\cite{rf:ogi99}, 
(especially if the hole is slowly rotating) so that 
far more power is extracted from the inner disk, or, more plausibly 
the region between the inner disk and the horizon than from the hole directly.
I emphasize the word directly because when energy is removed from 
gas in the ergosphere, the torques force that gas onto a lower energy,
perhaps even a negative energy orbit, than it would ordinarily follow. 
This means that when the mass is eventually captured by the 
hole, its gravitational mass increases by less than it would
do so in the absence of magnetic stresses. (It may even decrease.) 
We can therefore think of the energy as having been derived from the
spin of the hole even when the torque is applied to accreting gas.

One way to concentrate much more flux onto the horizon
of a rapidly spinning hole is to replace a thin disk with a thick toroidal
flow, such as is thought likely to develop when the gas does not
radiate effectively\cite{rf:ree82}\cite{rf:mei01}. 
It may even be necessary that 
this happen in order to form ultrarelativistic jets as this will 
minimize the effects of radiative drag.  
\subsubsection{Development of the Jet}
I have argued that a relativistic jet begins life 
close to a spinning black hole in an essentially purely 
electromagnetic form. However, what we observe are high energy
Compton-scattered gamma rays and lower frequency
synchrotron radiation. Given the large
electric fields present in the jets, it is very easy to imagine
pairs being created copiously near the hole. Now,
the pair energy density is unlikely to become very large here 
because it will be limited by annihilation.  Further from the hole,
though, when the energy density diminishes, pairs and gamma rays will carry a
progressively  larger fraction of the energy flux. 
\subsubsection{Gamma-ray Bursts}
As more is being learned about gamma ray bursts, the more they appear
similar to AGN. The inferred bulk Lorentz factors of GRBs 
and AGN are separated by less than an order of magnitude. GRBs are 
argued to have a low baryon fraction just like radio jets. Studies
of jets increasingly show them to be episodic phenomena as opposed to 
continuous flows. Finally, the appearance of achromatic breaks in the 
development of GRB afterglows has been interpreted as indicating that they 
too are jet flows beamed towards us, (though these 
observations may also be associated with the trans-relativistic 
evolution of spherical blast waves). If GRBs are mostly jets, then this 
reduces the energy per burst by two or three orders of magnitude
at the expense of increasing their overall frequency.

Many models of GRBs also  
involve the extraction of large amounts of energy from 
stellar mass black holes and electromagnetic processes occuring at 
or near the horizon of the black hole.
The magnetic fields and EMFs involved ($\sim10^{16}$~G,
$\sim10^{23}$~V) are much larger than those
associated with AGN but this should create no difficulty of principle. 
\subsection{Jet-Outflow Connection}
We can now complete the cycle and consider the interaction of the jet with 
the less collimated and slower outflow associated with the disk. The most 
important feature of this interaction is, of course, that the wind may 
be responsible for collimating the jet. It may provide an effectively 
invisible sheath that protects the rather fragile jet outflow 
from interaction with its environment. (This is one place where 
AGN and galactic superluminals may differ from GRBs. If GRBs really are 
strongly beamed, then this will probably have to be attributed to collimation
by a dense surrounding gas distribution, quite possibly a star. It is 
unlikely that any outflow that develops before the explosion that forms
the GRB will have enough pressure or inertia to provide any
collimation.  

A second feature of the jet-outflow interaction is that there will surely
be some entrainment of the electron-ion plasma and this should 
ultimately show up in the polarization observations which can, in principle
distinguish a pair plasma from a protonic plasma. Entrainment will 
be promoted by linear instabilities that can grow at the 
jet surface\cite{rf:app93}. 

Thirdly, there can be an exchange of linear
momentum with the surrounding outflow, even if there is minimal exchange 
of mass. This can, in turn, have two consequences. The jet, itself, 
is likely to develop a velocity profile so that different parts move with 
with different Lorentz factors. This implies
that a radio observer is likely to infer a value for the 
Lorentz factor from the rate of superluminal 
expansion that depends upon the inclination $i$ of the line of sight to the 
jet axis. Making the simplest assumptions, we expect that the 
measured Lorentz factor will be $\sim i^{-1}$. This could 
lead to some very strong biases in interpreting the statistical
properties of a sample of compact radio sources.  

These features could account for the difference between
Fanaroff-Riley Type I and Type II sources. Perhaps the former arise
when the outflow has more linear momentum than the jet which is 
ultimately decelerated to speeds of a few hundred km s$^{-1}$,
comparable with that of the outflow emanating from the outer disk.
Conversely, if the jet carries more linear momentum, it can drag along 
the outflow as presumably happens in the Type II sources, with little 
in the way of observed consequences.   
\section{Numerical Simulations} 
Much of what I have summarized is qualitative and conjectural and the 
debates that I have highlighted revolve largely around different prejudices
as to how magnetized, three dimensional flows behave in strong 
gravitational fields. There are serious issues of theory that need to be 
settled independent of what guidance we get from observations of 
astrophysical black holes and those of other sources, like the
solar corona, where magnetic field holds sway. (I might also remark 
that there is the strong prospect of laboratory experiment 
being highly relevant in teaching us the true laws of MHD.)

The best prospects probably lie with performing numerical 
simulations or experiments. Numerical MHD is coming of age
\cite{rf:mil00}\cite{rf:koi01}\cite{rf:haw01}\cite{rf:mei01a}. The 
sophistication of the difference schemes and visualization 
techniques is growing apace with the speed and especially the memory 
of large parallel-processing computers. Well-resolved 3-D (and even 4-D)
simulations are becoming increasingly common and they rarely fail to surprize
us. The symmetry-breaking involved in transitioning from 2-D to 3-D is
crucial and leads to qualitatively new phenomena. 

The key to using simulations productively is to isolate 
questions that can realistically be addressed
and where we do not know what the outcome 
will be and then to analyze the simulations so that we can learn what is
the correct way to think about the problem and to describe it in terms 
of eleemntary principles. Simulations in which 
the input physics is so circumscribed that they merely illustrate
existing prejudice are of less value! 
 
Let me frame some of these issues using a series
of questions, many of which are already being
tackled, as we have heard at this meeting.
\begin{itemize}
\item {\it What is the global evolution of the magnetorotational instability?}
The re-discovered linear instability is so important that it is irrelevant!
The non-linear evolution has been followed in 2- and 3-D, mostly
in shearing boxes, particularly with regard to following
the development of stress. 
Is there a cascade of wave energy to small scale through an inertial range
turbulence spectrum or does most of the energy follow an 
inverse cascade to form large loops as has also been 
conjectured? Another question is``What is the role of 
buoyant escape of magnetic
flux as well as the negative buoyancy associated with magnetic
tension as described above?''. Existing simulations suggest that it is too slow to be important. 
However only a few regimes have been studied. Most important of all,
we would like to know if large scale magnetic fields develop 
or can be sustained in Keplerian disks. One approach to this problem is to 
use simulations performed for a few dynamical time scales to try to measure 
the effective average resistivity for large scale fields. Another 
approach is look for signs of an inverse cascade developing in the
presence of  
strong differential rotation.
\item{\it What is the nature of the dissipation in magnetized flows?}
It is also necessary to understand the flow of
energy and, in particular, the nature of the dissipation, 
seen from a fluid perspective. This is a pre-requisite
to understanding dissipation from a kinetic perspective which, as 
the above makes clear, is crucial to understanding the radiative 
and consequently the dynamical properties of adiabatic flows. The most common
assumption is that the dynamical magnetic field creates a magnetosonic
wave spectrum that dissipates on some inner scale, small compared with the 
thickness of the disk. However it is also possible
that magnetic reconnection contributes to the dissipation 
in addition to being a switch to bring about
topological re-arrangement of the field. If reconnection heating
is significant, then is it the 
occasional large events that dominate the heating or is it the ongoing
``nanoflares'' that are most important. (This is an issue of active debate
in studies of solar coronal heating.)  
\item{\it How do magnetic fields develop in radiation-dominated disks?}
One interesting possibility is that they are essential in allowing a 
super-Eddington radiation flux to escape without blowing away the gas
by creating large density inhomogeneities.
Understanding the strength of the torques in the region where most of the 
energy is released is of central importance in modeling quasar spectra.
\item{\it How and where are disk coronae heated?}
X-ray observations of Seyfert galaxies are commonly and fairly convincingly
interpreted in terms of a model in which a coronal source, in the form
of a Comptonized power law extending up to $\sim100$~keV illuminates
a disk. The reflection spectra create the curved continuum and the fluorescent
iron lines. It has generally been supposed that the source is powered 
magnetically. However, it is again not clear if this is due to
a few large flares
at high altitude or a multitude of small flares occuring close to the disk.
In this case the sun may not be a very good guide, because accretion disks are
fast rotators in the sense that centrifugal force is more important 
than pressure gradients whereas the rotation that ultimately drives 
solar flares is only a small perturbation. Understanding the origin of the 
X-ray power law continua is also crucial for assessing if it will ever
be possible to perform reverberation mapping at X-ray energies.
\item{\it If large scale fields develop in accretion disks, how strong are 
the outflows and which mechanism dominates their formation?}
Several possibilities have been discussed, in addition to 
centrifugally-driven outflows. These include thermally-driven winds, outflows 
driven by the pressure of a strongly coiled magnetic field and outflows in 
which the field is basically poloidal. A major issue is the 
stability of these outflows and here, again, three dimensional simulations
make all the difference. Preliminary studies of centrifugal 
winds show that they are surprisingly resilient.
\item{\it Can electromagnetic power really be extracted from a 
spinning black hole and is it enough to power relativistic jets?}
Here there are two questions. The first is one of physics principle, as I have
discussed. What is probably necessary to settle the dispute alluded 
to above is to perform 
a time-dependent calculation in a Kerr metric. I have argued that 
it is sufficient to perform this calculation in the electromagnetic limit; 
there is no need to carry the unwanted baggage of inertial terms
in the limit that these are very small as must be the case if 
it is intended to account for ultrarelativistic jets. The second 
question is how much flux can be concentrated onto the black hole relative
to that which threads the inner disk? This must be a hydromagnetic 
calculation. 
\item{\it How do jets propagate, collimate and dissipate?}
There has been steady progress in describing disk winds numerically. What is
really needed now is to produce hybrid simulations that combine a 
ultrarelativistic jet core with a subrelativistic wind. The flows may 
become unstable, especially as toroidal field starts to dominate, but this
does not imply that jets destroy themselves catastrophically. Indeed many
of the maps from VLBI monitoring programs look like helically
unstable flows. 
\item{\it Can ultrarelativistic jets really be formed inside stars?}
As of this writing, the collapsar model appears to be the leading 
candidate for explaining the long duration GRBs. This is not on the 
face of it the most propitious environment to create an ultrarelativistic,
baryons-starved 
jet. I have argued that electromagnetic 
effects, similar to those already invoked for AGN 
jets seem to be the most promising way to create the enormous powers 
required. What needs to be demonstrated is that the outflow is not 
``poisoned'' by baryons by the time that it reaches the surface of the star.
It is not, in my view necessary to collimate the jet very tightly
or achieve high bulk Lorentz factor as the flow leaves the stellar surface.
As long as the the emergent flow has a high enthalpy per baryon,
it will expand on a Mach cone and achieve its high terminal 
speed some distance from the star. This, incidentally, provides a natural 
explanation for the otherwise puzzling initial 
absence of causal contact across
an expanding ultrarelativistic jet - similar to the famous 
``hello-goodbye-hello'' behavior of an inflationary universe. 
\end{itemize}  

It is easier to make such lists than to implement them.
\section*{Acknowledgements}
I thank the Mayor of Nishinomiya 
City and Fumio Takahara  for their gracious hospitality and the latter
for organizing a lively symposium and his patience.
I am indebted to Mitch Begelman, Shinje Koide, Dave Meier,
Ralph Pudritz and Brian
Punsly for stimulating discussions (even if they do not agree with 
all of the above). I acknowledge support under NASA grant 5-2837.

\end{document}